\documentclass[aps,preprintnumbers,showpacs,twocolumn,superscriptaddress,floatfix]{revtex4}

\usepackage{amsmath}
\usepackage{latexsym}
\usepackage{graphicx, color}
\usepackage{epstopdf}           
\usepackage{times}

\def\be{\begin{equation}}
\def\ee{\end{equation}}
\def\bea{\begin{eqnarray}}
\def\eea{\end{eqnarray}}
\def\scri{\mathcal{J}^+}

\begin{document}

\title{Unambiguous determination of gravitational waveforms from 
       binary black hole mergers}


\author{C. Reisswig}
\affiliation{
Max-Planck-Institut f\" ur Gravitationsphysik, Albert-Einstein-Institut,
14476 Golm, Germany
}

\author{N.~T. Bishop}
\affiliation{
Department of Mathematics, Rhodes University, Grahamstown 6140, South
Africa
}
\affiliation{
Department of Mathematical Sciences,
University of South Africa, Unisa 0003, South Africa
}

\author{D. Pollney}
\affiliation{
Max-Planck-Institut f\" ur Gravitationsphysik, Albert-Einstein-Institut,
14476 Golm, Germany
}

\author{B. Szil\'{a}gyi}
\affiliation{
Theoretical Astrophysics, California Institute of Technology,
Pasadena, CA 91125, USA
}

\begin{abstract}
Gravitational radiation is properly defined only at future
null infinity ($\scri$), but in practice it is estimated
from data calculated at a finite radius. We have used characteristic
extraction to calculate gravitational radiation at $\scri$ for the
inspiral and merger of two equal
mass non-spinning black holes. Thus we have determined the
first unambiguous merger waveforms for this problem. The
implementation is general purpose, and can be
applied to calculate the gravitational radiation, at $\scri$, given
data at a finite radius calculated in another computation.
\end{abstract}


\maketitle

{\it Introductory Remarks.} The computation of gravitational radiation
from black hole merger events has attracted considerable attention,
since the pioneering work by Smarr and
collaborators~\cite{Smarr75, Smarr76, Smarr78b}.  With the advent of
ground-based laser interferometric gravitational wave detectors, as
well as the prospect of the Laser Interferometer Space Antenna (LISA),
interest in the problem has considerably increased.  The measurement
of gravitational waves will soon provide an important probe of
strong-field nonlinear gravity, the domain of many fundamental
questions in astrophysics.  The sensitivity of LISA, and of the
upcoming advanced ground-based detectors AdLIGO and AdVirgo, is so high
that even an error in the waveform calculation of $0.1\%$ (in a sense
made precise later) could lead to an incorrect interpretation of
the astrophysical properties of a source, or of a test of
general relativity.  Nowadays, there are
several codes that can produce a stable and convergent simulation of a
black hole spacetime. However, a particular difficulty with measuring
gravitational radiation arises from the fact that, in general relativity, it
cannot be defined locally but is defined only at future null infinity
($\scri$), which, physically, is the limit that is approached by radiation
moving at the speed of light away from an isolated source. Since numerical
evolutions are normally carried out on finite domains, there is a
systematic error caused by estimating the gravitational radiation from
fields on a worldtube at finite radius and the uncertainty in how
it relates to measurement at $\scri$ \cite{Kocsis:2007}.  
Even if this error is small,
the expected sensitivity of AdLIGO, AdVirgo and LISA implies that it
is important to obtain an accurate result.

A rigorous formalism for the global measurement of gravitational
energy at null infinity has been in place since the pioneering work of
Bondi, Penrose and collaborators in the 1960s~\cite{Bondi62,
Penrose:1963}; and subsequently, techniques for calculating gravitational 
radiation at $\scri$ have been developed. The idea which we pursue here
is to combine a Cauchy or ``3~+~1'' numerical relativity code with a
characteristic code~\cite{Bishop93}. Given astrophysical initial data,
such a method has only discretization error~\cite{Bishop96}, and a
complete mathematical specification has been
developed~\cite{Bishop98b}. There have been efforts to implement this
method, often called Cauchy characteristic extraction (CCE), or
characteristic extraction~\cite{Babiuc:2005pg,Babiuc:2009}. Previous
work has considered test problems rather than that of
the inspiral and merger of two black
holes or neutron stars. Also, earlier efforts have combined the Cauchy
and characteristic algorithms within the same code.

Here, we describe the implementation of a CCE
code as well as results
obtained from the code for the inspiral and merger waveform of two
equal mass, non-spinning, black holes. The waveforms are calculated at
$\scri$, and are thus the first unambiguous waveforms which have been
obtained for this problem, in the sense of being free of gauge 
or finite-radius effects. 
Further, the code is general purpose, in that it is
independent of the details of the Cauchy code, requiring only that it
prescribes the required geometrical data on a world-tube. Thus its
application to other astrophysical problems will be straightforward.

For the specific problem of a binary black hole (BBH) merger, we show
that the waveform obtained at $\scri$ contains only numerical error
and is gauge-invariant.  We demonstrate second-order
convergence to zero in the amplitude and phase differences between two
CCE runs using boundary data at different radii.  We
compare the waveform obtained at $\scri$ with a finite-radius
extrapolated waveform, and find that the corrections introduced by CCE
are visible in the groundbased detectors AdLIGO and AdVirgo, as well
as the space-based LISA detector.

{\it Cauchy Evolution.} The scenario we envision is an
isolated system (perturbed single body, or gravitationally bound
binary), in a region on which the Einstein (and possibly
hydrodynamic) equations must be solved. A standard procedure for doing
this is to formulate the equations as an initial-boundary-value, or
Cauchy, problem, in which data for the 3-metric and its embedding is
prescribed at a given time on a closed region of the spacetime,
$\Sigma_t$. These are evolved according to the Einstein equations on
the interior of the domain, and artificial conditions on the timelike
boundary, $\partial\Sigma_t$.  The first stable evolutions of a binary
black hole system were carried out by
Pretorius~\cite{Pretorius:2005gq}. Two approaches to the
evolution of the interior equations are in use: (a) the harmonic
formulation of the Einstein equations with excised black hole
interiors~\cite{Pretorius:2005gq, Scheel:2008rj}; (b) the BSSNOK 
(see \cite{Alcubierre:2008} and references therein)
evolution system with the black holes specified as moving
``punctures''~\cite{Baker:2005vv, Campanelli:2005dd}.

For the Cauchy evolutions used here, we have followed the latter
approach using the formulation
outlined in~\cite{pollney:2007ss}. The spacetime is discretised using
finite differences on Cartesian grids and Berger-Oliger mesh
refinement in the neighbourhood of the black
holes~\cite{Schnetter-etal-03b}. 
The wave zone is discretised by six
overlapping coordinate patches with spherical topology. Interior
boundary data between adjacent patches are communicated by
interpolation~\cite{Pollney:2009MP-unpublished-a, Pollney:2009MP-unpublished}.

The outer boundary condition on the exterior of the domain is given by
a linear outgoing wave condition on each of the evolved
tensor components. Importantly, through the use of spherical grids in the
wave zone, a sufficient resolution can be maintained even to a distant
outer boundary, reducing the effect of grid reflections common in
mesh-refinement codes. The size of the evolution domain is chosen
according to the amount of time required, $T$, and the location of the
outermost measurement sphere, $r_i$. Since physical as well as
constraint violating modes propagate with the speed of light, an outer
boundary located at $r_{\partial\Sigma_t} > T + r_i$ ensures that
measurements are causally disconnected from the influence of the outer
boundary.

\begin{figure}
\centering
\includegraphics[width=8.5cm,clip,trim=120 470 50 100]{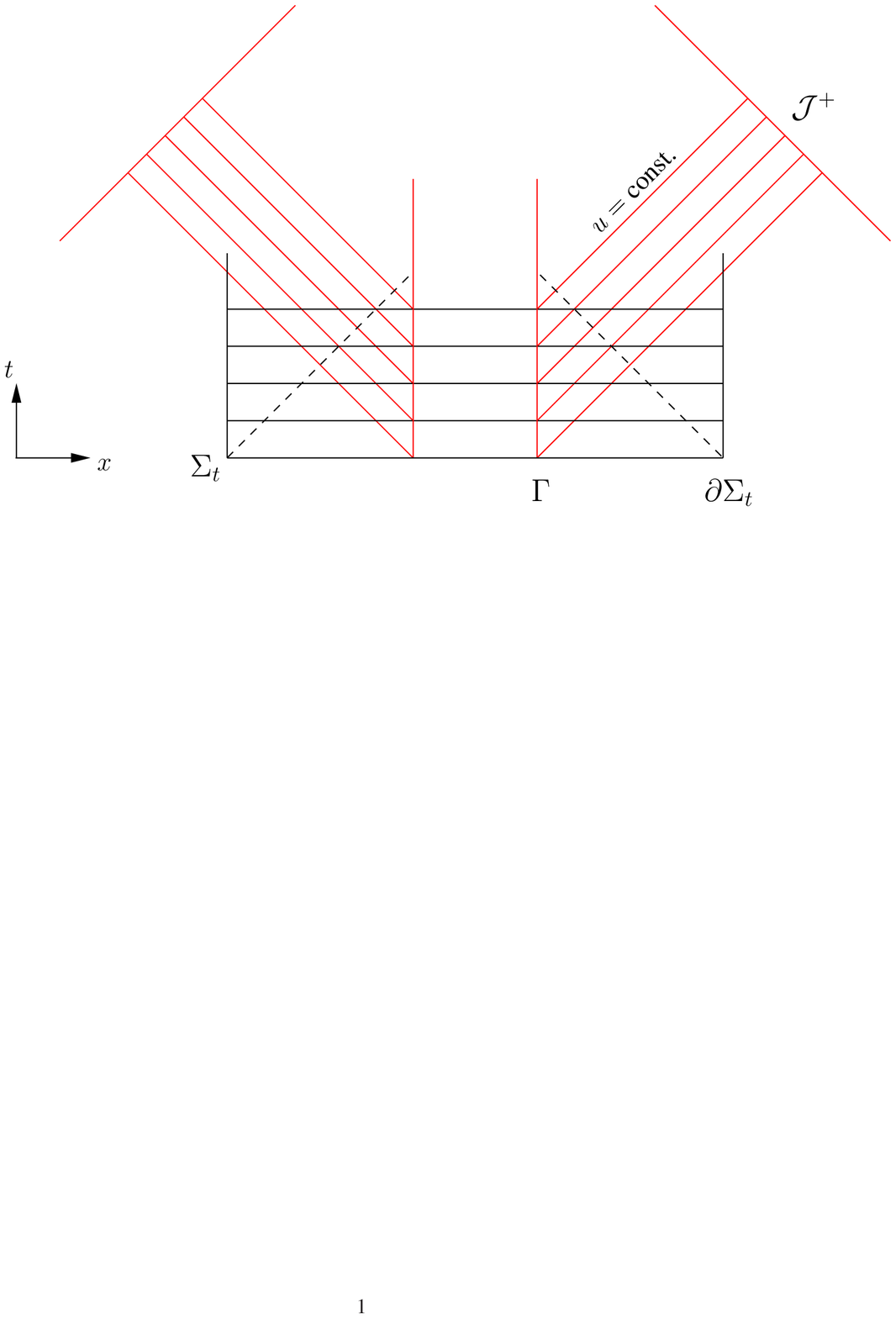}\hfill
\caption{
  Schematic of the CCE algorithm with two spatial dimensions suppressed.
  Spacelike slices, $\Sigma_t$ are
  evolved according to the Cauchy evolution scheme (horizontal lines).
  Geometrical data is recorded on a world-tube, $\Gamma$, which is
  used as interior boundary data for a characteristic evolution along
  $u=\text{constant}$ null surfaces, transporting the data to $\scri$.
  The outer boundary of the Cauchy domain, $\partial\Sigma_t$ is chosen
  so that it is causally disconnected from $\Gamma$ over the course of
  the evolution.
}
\label{fig:CCE}
\end{figure}

{\it Characteristic extraction.}  Our implementation of CCE
is based on the mathematical prescription given
in~\cite{Bishop98b}, and here we provide only an outline.
The process is illustrated schematically in
Fig.~\ref{fig:CCE}. Within a Cauchy simulation that uses
Minkowski-like coordinates $(t,x,y,z)$, we define a worldtube $\Gamma$
by $x^2+y^2+z^2=r^2_\Gamma$, and compute the lapse $\alpha$, the shift
$\beta^i$ and the 3-metric $\gamma_{ij}$ on $\Gamma$, as well as their
first time and radial derivatives.
This data is then decomposed into spherical harmonics. The Cauchy
code writes this spherical harmonic coefficient data to file, and
later the CCE code postprocesses the data to
reconstruct the 4-metric on the inner world tube. In this way, the
CCE code is general purpose, as it runs independently of
whatever Cauchy code was used to generate the worldtube data.

The CCE code defines angular
coordinates $\phi^A$ as well as a time coordinate $u$ (=$t$) on
$\Gamma$, and constructs outgoing null geodesics with affine parameter
$\lambda$. It then transforms the Cauchy 4-metric to
$(u,\lambda,\phi^A)$ coordinates, and calculates a surface area radial
coordinate $r_S$, making the coordinate transformation to
$(u,r_S,\phi^A)$ coordinates, in order to obtain the Bondi-Sachs
metric data in a neighbourhood of $\Gamma$. This provides the inner
boundary data for the characteristic evolution, using a
Cactus~\cite{Allen99a} implementation of the PITT null evolution code
with square stereographic coordinate patches~\cite{Bishop97b}.
The characteristic code uses coordinates based on
outgoing null cones, and so the equations remain regular when the
radial coordinate is compactified (by $r_S\rightarrow
z=r_S/(r_S+r_{S\Gamma})$), and in this way $\scri$ is included on the
computational grid. The code computes the gravitational radiation at
$\scri$ as the Weyl component $\psi_4$. The coordinate-independent quantity 
$\psi_4$ is commonly used in numerical relativity; in appropriate
coordinates, it is the second time derivative of the strain measured
by a detector.

{\it Binary black hole evolution.} We have carried out fully
relativistic evolutions of an equal-mass non-spinning binary black
hole inspiral and merger.  The initial data parameters for the closely
bound black hole are determined by performing a post-Newtonian
evolution from large separation in order to determine the momenta for
low-eccentricity (quasi-circular) trajectories~\cite{Husa:2007rh}. 
The subsequent full nonlinear numerical relativity evolution proceeds for
approximately 8 orbits ($1350M$), followed by the merger and ringdown 
lasting another $100M$.

The evolutions have been carried out at two different grid resolutions
in order to verify the convergence of the numerical scheme. The grid
settings for the Cauchy code are: The central Cartesian
grid consists of 6 levels of 2:1 mesh-refinement, with coarse grid
spacings of $h=0.96M$ and $h=0.64M$, respectively.  A grid of
spherical topology covers the far field, $r\in[35M,3600M]$; so that,
during the time period of interest, the outer boundary is causally
disconnected from any extraction sphere (see Fig.~\ref{fig:CCE}). The radial
spacing is commensurate with the coarse Cartesian grid at the
interface, and (for reasons of efficiency) is gradually scaled to
$h=3.84M$ and $h=2.56M$ at the outer boundary for the two runs. We use
a corresponding $N_{\rm ang}=21$ and $N_{\rm ang}=31$ points in each of the
angular directions per patch.

\begin{figure}
  \centering
  \includegraphics[width=8.5cm]{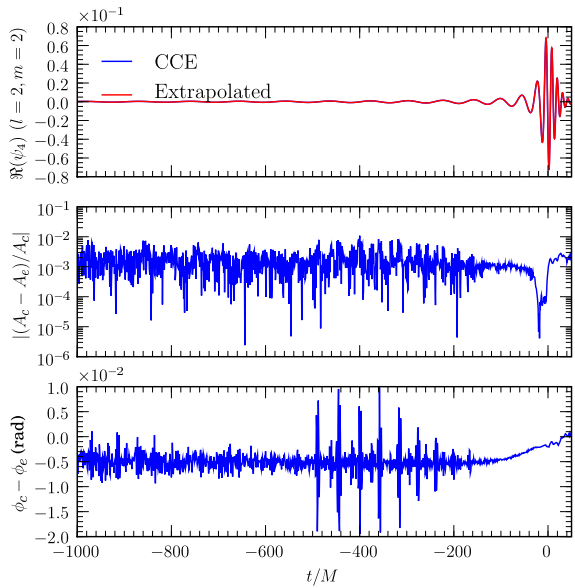}\hfill
  \vskip -2.5mm
  \caption{Inspiral, merger and ringdown phase of $\psi_4(\ell=2,m=2)$
    as obtained from finite radius extrapolation (red) and at $\scri$
    (blue). The waveforms are aligned at their peaks.
    There is a maximum difference of $1.08\%$ in the amplitude and a
    dephasing of $0.019$ radians between the two waves. These differences
    can introduce systematic errors to parameter estimation of events detected
    at high SNR.}
    \label{fig:waveform}
  \vskip -3mm
\end{figure}

Characteristic boundary data were
interpolated onto world-tubes located at $r=100M$ and
$r=200M$, and stored in the form of spherical harmonic coefficients,
up to $\ell=8$, which was found to be the highest resolved mode.
The resolutions of the characteristic evolutions are
set up according to the respective resolutions of the Cauchy
run.  We use $N_r=321$ and $N_r=481$ radial points, with
$N_{\rm ang}=51$ and $N_{\rm ang}=76$ angular points per angular patch.
The dominant $\ell=2$, $m=2$ mode of the gravitational waveform resulting
from the numerical evolution is plotted in Fig.~\ref{fig:waveform}, to
be described in more detail below.

\begin{figure}
  \centering
  \includegraphics[width=8.5cm]{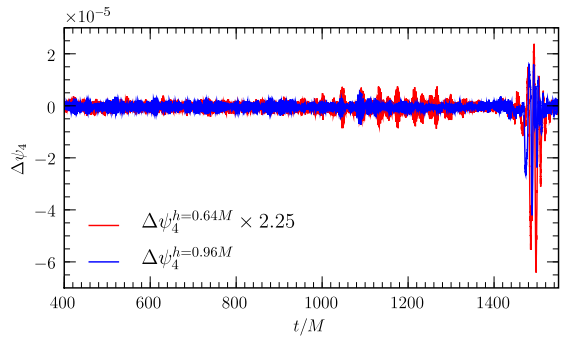}
  \caption{
    Differences $\Delta\psi_4$ in the amplitude of $\psi_4(\ell=2,m=2)$
    between two characteristic runs using boundary data from $R_\Gamma=100M$
    and $R_\Gamma=200M$. 
    The red curve shows the difference at resolution $h=0.96M$
    while the blue curve shows the difference for $h=0.64M$, scaled
    so as to line up for second-order convergence. The expected second
    order convergence of our code is thus demonstrated.
  }
  \vskip -2.5mm
  \label{fig:wave-diff}
\end{figure}

Invariance with respect to the world tube location is
demonstrated in Fig~\ref{fig:wave-diff}. We have
considered the differences between waveforms at $\scri$ resulting from
two independent characteristic evolutions using boundary data at
$r_{\Gamma}=100M$ and $r_{\Gamma}=200M$, respectively, and for two
resolutions, $h=0.96M$ and $h=0.64M$. The difference between the
results should be entirely due to the discretisation error, and indeed
this is what we find. The differences converge to zero with approximately
second-order accuracy, as expected for the null evolution code. The figure
displays the differences $\Delta\psi_4$ in the amplitude of the
wave mode $\psi_4(\ell=2,m=2)$ for resolution $h=0.96M$ and $h=0.64M$
scaled for second-order convergence. The same order of convergence is
also obtained for higher order modes such as $\psi_4(\ell=4,m=4)$ (not
displayed). 
The differences between the waveforms at $\scri$ for resolution
$h=0.64M$ are of order of $0.03\%$ in amplitude with a dephasing
of $0.002$ radians.

The CCE waves can be used to evaluate the quality of standard finite
radius measurements, extrapolated to $r\rightarrow\infty$; previously,
this was the most accurate option available.
We do so by
finding the Weyl component $\psi_4$ relative to
a radially oriented null tetrad~\cite{pollney:2007ss} (we prefer $\psi_4$ 
to gauge-invariant perturbative methods~\cite{Moncrief74, Zerilli70,
Nagar:2005ea}). We have
evaluated $\psi_4$ at six radii ($r=280,300,400,500,600,1000M$) and
extrapolated. Details are given in~\cite{Pollney:2009MP-unpublished},
and the error is estimated as $0.03\%$ in
amplitude and $0.003$ radians in phase.

In Fig.~\ref{fig:waveform}, we compare the extrapolated waveform
with that calculated at $\scri$ via CCE.  The differences
between the two waveforms have maximum and mean values of $1.08\%$ and
$0.166\%$ in amplitude, and $-0.019$ and $-0.004$ radians
in phase, respectively. 
That is,
for the resolutions used, the numerical error in the characteristic
evolution (see Fig.~\ref{fig:wave-diff}) is smaller by one order of
magnitude than the error between the extrapolated and characteristic
waveforms in both amplitude and phase.  Further, we note that the
estimated error in the extrapolation is itself much smaller than the
actual error between characteristic waveform and extrapolated
waveform. This indicates that the systematic error in extrapolation has,
previously, been underestimated. The correction is towards slightly
larger amplitudes and frequencies. CCE post-processes data produced by a
Cauchy code, and as such there is an additional cost. However, it is
relatively small: the Cauchy run reported here took $\sim336$ hours, and then CCE
required $\sim10$ hours.

Will the small correction to waveforms introduced by CCE be relevant
to interpreting observational data? The answer will depend on the
signal-to-noise ratio (SNR) of the event. At low SNR, whether CCE or
extrapolated waveforms are used as a template will not affect physical
interpretation. This is particularly relevant, as
numerical waveforms are being constructed with the intention of
evaluating and parametrising detector templates and search
algorithms~\cite{Aylott:2009ya}, and to constrain analytic
models~\cite{Buonanno:2009qa, Damour:2009kr, Ajith:2007kx:longal, Ajith:2009bn}.
Our results indicate that extrapolations from a finite
radius can be used to construct detector templates well within the
accuracy standards required by matched filtering algorithms.

However, at large SNR, the differences are significant
to the determination of the physical parameters
of a model measured in detector data.
To demonstrate this, we follow
methods described in \cite{Lindblom:2008cm, Hannam:2009hh_pbl} to determine the minimum
SNR needed for a detected signal from a merger event
to lead to different parameter estimates depending on which waveform is used
as a template. Table~\ref{table-snr} displays
the results for selected masses, indicating the maximum distance at which
the difference between
the waveforms will be relevant for the given merger event.

The difference between the waveforms is unlikely to be relevant for
LIGO, (e)LIGO and Virgo. 
Reasonable stellar mass black hole merger
rates are expected only
for a volume encompassing
sources up to a distance of at least 100Mpc. Thus, there
may well be events detected by AdLIGO and AdVirgo 
for which the difference
is important.  Finally, the differences will certainly be relevant for
LISA as they will be applicable to any supermassive black hole
merger event throughout the visible universe ($cH^{-1}$ is the Hubble
radius).

\begin{table}
  \begin{tabular}{l|ll}\hline\hline
    Detector & Masses & Maximum distance \\ \hline
    LIGO & $50M_{\odot}+50M_{\odot}$ & 5Mpc \\
    (e)LIGO & $50M_{\odot}+50M_{\odot}$& 8Mpc \\
    Virgo & $50M_{\odot}+50M_{\odot}$& 14Mpc \\
    AdLIGO & $50M_{\odot}+50M_{\odot}$& 197Mpc \\
    AdVirgo & $50M_{\odot}+50M_{\odot}$& 177Mpc \\
    LISA & $10^7M_{\odot}+10^7M_{\odot}$& $>cH^{-1}$\\ \hline\hline
  \end{tabular}
  \caption{Maximum distance at which the difference between the
    extrapolated waveform and that at $\scri$ would be significant for a
    black hole merger event.}
  \vskip -3mm
  \label{table-snr}
\end{table}

{\bf Acknowledgments.} 
The authors thank
Stanislav Babak, Luciano Rezzolla and Jeffrey Winicour for their helpful input.
CR, DP and BS thank the University of South Africa, and NTB thanks
Max-Planck-Institut f\" ur Gravitationsphysik, for hospitality.
This work was supported by
National Research Foundation, South Africa; Budesministerium f\"ur Bildung und
Forschung, Germany; and DFG grant
SFB/Transregio~7 ``Gravitational Wave Astronomy''.
DP was supported by a grant from the VESF.
BS was supported by grants from the Sherman Fairchild Foundation, by NSF grants
DMS-0553302, PHY-0601459, PHY-0652995, and by NASA grant NNX09AF97G.
Computations were performed at the AEI, at LRZ-M\"unchen, on Teragrid clusters
(allocation TG-MCA02N014), and LONI resources at LSU.

\bibliography{aeireferences}

\end{document}